# Growth of Pure and Intercalated ZrTe$_2$, TiTe$_2$ and HfTe$_2$ Dichalcogenide Single Crystals by Isothermal Chemical Vapor Transport.

Lucas E. Correa[1], Leandro R. de Faria[1], Rennan S. Cardoso[1], Nabil Chaia[2], Mario S. da Luz[3], Milton S. Torikachvili[4] and Antonio J. S. Machado[1]

[1]Escola de Engenharia de Lorena, University of Sao Paulo, P. O. Box 116, Lorena, 12602-810, Brazil
[2]Instituto de Ciência e Tecnologia, Universidade Federal de Alfenas, Rod. José Aurélio Varela, 11999, 37715-400 Poços de Caldas, Minas Gerais, Brazil
[3]Instituto de Ciências Tecnológicas e Exatas, Universidade Federal do Triângulo Mineiro, Uberaba, Minas Gerais, Brasil
[4]Department of Physics, San Diego State University, San Diego, California 92182-1233, USA

## ABSTRACT

We report on a modified chemical vapor transport (CVT) methodology for the growth of pure and intercalated Zr, Ti, and Hf dichalcogenide single crystals, e.g. ZrTe$_2$, Gd$_{0.05}$ZrTe$_2$, HfTe$_2$, and Cu$_{0.05}$TiTe$_2$. While the most common method for CVT growth is carried out in quartz tubes subjected to a temperature gradient between the charge and the growth location, the growth using this isothermal-CVT (ICVT) method takes place isothermally in sealed quartz tubes placed horizontally in box furnaces, using iodine (I$_2$) as the transport agent. The structure and composition of crystals were determined by means of X-ray diffraction (XRD), scanning electron microscopy (SEM), and induced coupling plasma (ICP). The crystals grown with this method can be large, and show excellent crystallinity and homogeneity. Their morphology is plate-like, and the larger dimensions can be as long as 15 mm.

## Introduction

The study of the intrinsic physical properties of materials depends largely on the availability of high-quality single crystals. While high-quality polycrystalline specimens can be good enough for many experiments, addressing anisotropic behaviors or properties where grain boundaries interfere requires single crystals.

There are a number of techniques for synthesizing single crystalline materials. The techniques most often utilized to grow crystals of intermetallic compounds are chemical vapor transport (CVT) [1-4], Czochralski method [5-8], flux growth [9-11], and Bridgman methodology [12-14]. The choice depends largely on a careful examination of the phase diagram, chemical and physical properties of the constituents, and of course, available instrumentation. The conventional CVT technique relies on the transport of the gas phase of two or more dilute species using a temperature gradient as the driving force between the source and the sink, which makes the method quite versatile. The successful transport of elements in the gaseous phase and growth of single crystals depend largely on the thermodynamic characteristics of the system. CVT is well known to produce single crystals of elevated purity and excellent crystallinity. The most often transport agents used to form a vapor phase of the constituents are halogens elements or compounds [15,16]. Halides species of the elements to be transported are formed at the source, migrate over the temperature gradient, and decompose as result of a reversible reaction according to the equation 1.

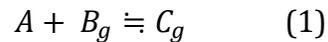

$$A + B_g \leftrightharpoons C_g \qquad (1)$$

where A is a solid or liquid phase, $B_g$ is a gaseous substance (halogens elements or compound), and $C_g$ is a gaseous product [17]. A typical example is the growth of $Cr_3Ge$ using iodine as the transport agent, where A is polycrystalline $Cr_3Ge$, $B_g$ is iodine, and $Cr_3Ge$ are single crystals grown at 880 °C upon transportation from 780 °C [18]. The solid phase A becomes mobile as it gets volatilized by the transport agent $B_g$ and incorporated in $C_g$. The result of the transport of the A phase to the sink yields a seed specimen that can grow steadily, resulting in high quality single crystal. The single crystal growth may occur in the hot or cold zone, depending on whether the formation reaction is exothermic or endothermic, respectively.

A common feature for CVT, flux, Bridgman, Czochralski, and other growth methodologies is that the driving force for mass transport and growth is a temperature gradient. In this work we show that pure and intercalated single crystals can be grown using CVT under isothermal conditions, where the driving force is not a temperature gradient, but the chemical potential gradient between the gas phase and the stoichiometric composition. The isothermal method yielded pure and intercalated single crystals of $ZrTe_2$, $HfTe_2$ and $TiTe_2$ having high purity and crystallinity. The intercalating species included most 3d-, 4d-, and 5d- metals, as well as most lanthanides.

## Experimental Procedure

Polycrystalline precursors of $RTe_2$, $T_xRTe_2$ and $L_xRTe_2$ ($T$ = transition metal, e.g. Co and Ni; $R$ = refractory metal, e.g. Zr, Hf and Ti; and $L$ = Gd, Ce and Dy) were synthetized via conventional powder metallurgy, using high purity powders of the constituents. The nominal composition of the intercalating metals (Co, Ni, Ce, Gd, Cu and Dy) was kept at x = 0.1. Stoichiometric amounts of the powders for each sample were ground thoroughly in an agate mortar and pressed into pellets. The pellets were encapsulated in quartz tubes, sealed under a pressure of ≈ 100 mTorr, and reacted at 950 – 1000 °C for 48 hours. The pellets were subsequently reground, pressed, encapsulated, and heat-treated again at the same temperature for additional 24 hours. This procedure was important for the formation of the compound as well as to ascertain that intercalation had been accomplished. The resulting polycrystalline materials were pelletized and encapsulated together with a small amount of iodine. The amount of iodine was calculated for each ampoule and had the approximate values of 0.6, 1.3, and 2.5 mg/cm³, yielding total pressures of ≈ 0.25, 0.5 and 1 atm in the quartz tubes, respectively. The tubes were kept at 900 °C (or 1000 °C) for 10 days, after which they were quenched in ice water. The resulting single crystals grew out of the pellets (Fig. 1) and were gently removed using tweezers. The composition of all crystals was analyzed by means of energy dispersive spectroscopy (EDS) ancillary to a Hitachi TM 3000 scanning electron microscope (SEM). The composition of a subset of these single crystals was also determined by induced coupling plasma (ICP), yielding good agreement with the composition determined from EDS. The crystallographic structure of the crystals was determined by X-ray diffraction using a PANalytical diffractometer (model Empyrean), equipped with a texture goniometer.

**Results and Discussion**

Successful ICVT growth from a ditelluride precursor yielded single crystals as large as $\approx 15 \times 8 \times 0.2$ mm$^3$, as shown in Fig. 1. This size is typical of the largest crystals in each batch, whether they were pure of intercalated with a transition metal, or a lanthanide element. While our discussion is valid for all group IVB tellurides (intercalated or not), our following discussion will focus mainly on ZrTe$_2$-based samples.

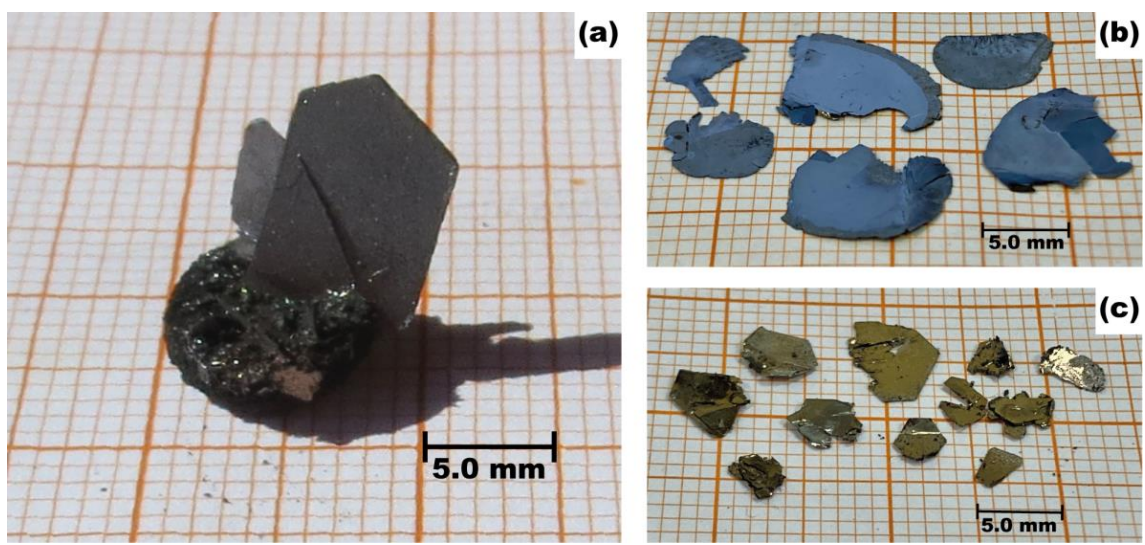

*Figure 1 – Single crystalline (a) Ce$_{0.04}$ZrTe$_2$, (b) Cu$_{0.05}$TiTe$_2$ and (c) HfTe$_2$ grown out of polycrystalline precursor pellets using the ICVT method. The longest dimension is $\approx$ 15.0 mm. This picture is representative of all crystals grown in this study.*

The driving force for usual CVT growth is a temperature gradient, which promotes mass transport between two slightly different thermodynamic conditions. However, when the transport agent reacts with the polycrystalline precursor a small chemical potential gradient is created, which may be sufficient to generate mass transport in the absence of a temperature gradient, leading to isothermal growth. The chemical potential is defined as the Gibbs free energy normalized to the elemental concentration, and is given by:

$$\mu_A = \left(\frac{\partial G}{\partial n_A}\right)_{T,P} \qquad (2)$$

where $\mu_A$ and $n_A$ are the chemical potential and the number of moles of substance A, respectively. Considering a reaction in a binary system as an example, e.g. A + B, under thermodynamic equilibrium we will have:

$$d\mu_A n_A + d\mu_B n_B = 0, \quad (3)$$

which results in:

$$d\mu_A = -\frac{n_B}{n_A} d\mu_B \quad (4)$$

The Zr-Te binary phase diagram [Ref. 19] suggests that the $ZrTe_2$ structure is stable over a wide homogeneity range, i.e., the Te content can vary between 1.6-2.0. A close investigation of Eq. 2 suggests that variations in the chemical composition of the compound induce modifications of the chemical potential of the species. It is tempting to consider that at high temperatures, the reaction of the iodine with the outer layers of the grains generates an activity gradient from the center of the grains towards the interface with the gas phase. Iodine then forms a complex volatile substance on the surface of the granules, effectively peeling out this outer layer, and therefore leaving a slightly modified stoichiometry behind. Because of the gradient of the chemical species through the grains and possibly the pellets, thermodynamic equilibrium between the substrate and the gas phase cannot be reached. A slight mismatch between the chemical potentials of the pellet and the gaseous phase around it create the necessary conditions for mass transport from the bulk to the surface. Once the concentration of the gas phase in the quartz tube reaches saturation, and the chemical potential of the species in the remaining pellet is lower than that of the species in the gas phase, crystal growth starts to take place by inversion of the gas flux. This growth process continues until saturation can no longer be sustained, given that the formation of crystals starts impoverishing the solution. As shown in Fig. 2, the nucleation of crystal seeds can be observed by naked eye on the surface of the pellets within 24 hours of isothermal conditions. Larger crystals start developing from the seeds, and it takes about 7 days to reach the largest sizes (Fig. 1)

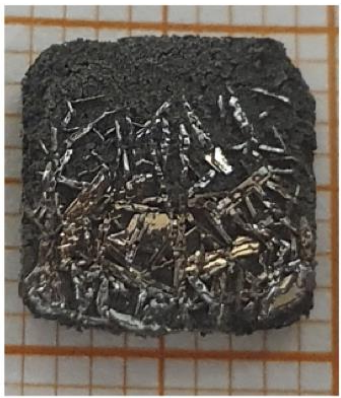

*Figure 2 –Image of the surface of a $Dy_{0.1}ZrTe_2$ pellet after 24 hs at 1000 °C. The small reflective fragments are nucleation centers.*

It is important to notice that the large size of the crystals shown is quite typical of the ICVT growth. Given the slight off-stoichiometry of the intercalated compositions, the growth condition remains favorable for a long time, i.e., until the solution becomes depleted.

In order to probe the relevance of the amount of transport agent to the ICVT growth, in this case iodine, we tested growth under three concentrations, 0.6, 1.3 and 2.5 g/cm³. Crystals yielded under a 0.6 g/cm³ iodine concentration had an average size of 4.0 mm, while the 1.3 g/cm³ iodine concentration yielded average crystal sizes in the 15-17 mm range, and the 2.5 g/cm³ concentration yielded smaller average sizes of about 7.0 mm, suggesting that some optimization is necessary in order to control the number of seeds and growth sizes. Table I summarizes these results.

Table I –Average size of the grown single crystals as a function of the iodine concentration used in the isothermal CVT growth.

| Iodine Concentration (g/cm³) | Average size (mm) |
|:---:|:---:|
| 0.6 | 4.0 |
| 1.3 | 17.0 |
| 2.5 | 7.0 |

The growth data under the optimized iodine concentration of 1.3 g/cm³ reveal that the larger dimension of the crystals starts to saturate after about 7 days of isothermal growth, as indicated in Fig. 3 for $Dy_xZrTe_2$ crystals.

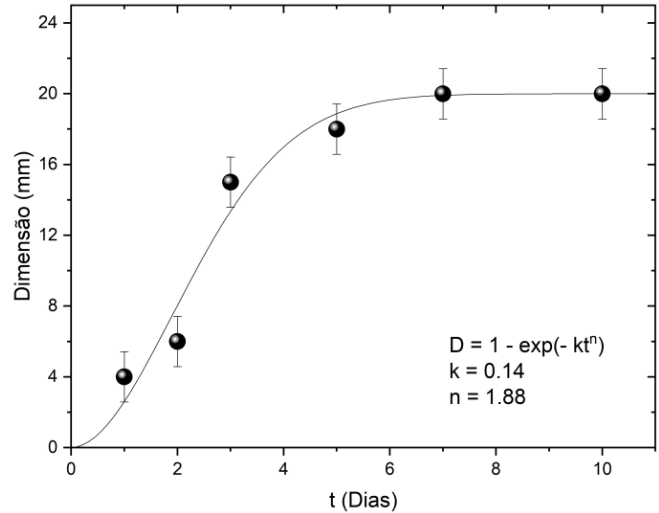

*Figure 3 –Largest dimension of* $Dy_xZrTe_2$ *crystals as a function of isothermal growth (T = 1000 °C), using 1.3 g/cm³ of iodine as the transport agent.*

The average crystal size is linked to the fraction of the gasification that takes place during the growth, and therefore it can be modeled with the Kolmogorov–Johnson–Mehl–Avrami (KJMA) expression for isothermal transformation kinetics [20], which is given by Eq. 5, known as the Avrami equation:

$$y = 1 - \exp{(-kt^n)} \qquad\qquad \text{Eq. (5)}$$

where y represents the fraction of the pellet sample converted to single crystals, t is the transformation time, and $k$ and $n$ are fit parameters. The $k$ parameter is correlated with seed nucleation and growth rates. The Avrami coefficient $n$ is linked to the dimensionality of the crystals [21]. The fit of the growth data for the $Dy_xZrTe_2$ single crystals of this work yielded n ≈ 1.88, consistent with the 2.0 value expected for growth of primarily two-dimensional materials [22]. This type of behavior has been observed also in intercalated $TiTe_2$- and $HfTe_2$-based materials, both of which with $CdI_2$-type structure. It is important to notice that the compounds crystallizing with this prototype structure are regarded as *van der Waals* compounds.

The typical high quality of the crystals grown using this method is supported by the θ-2θ X-ray diffraction scans with the incident beam directed only to the flat surface of the crystals, as shown in Fig. 4 for $Ni_{0.05}ZrTe_2$.

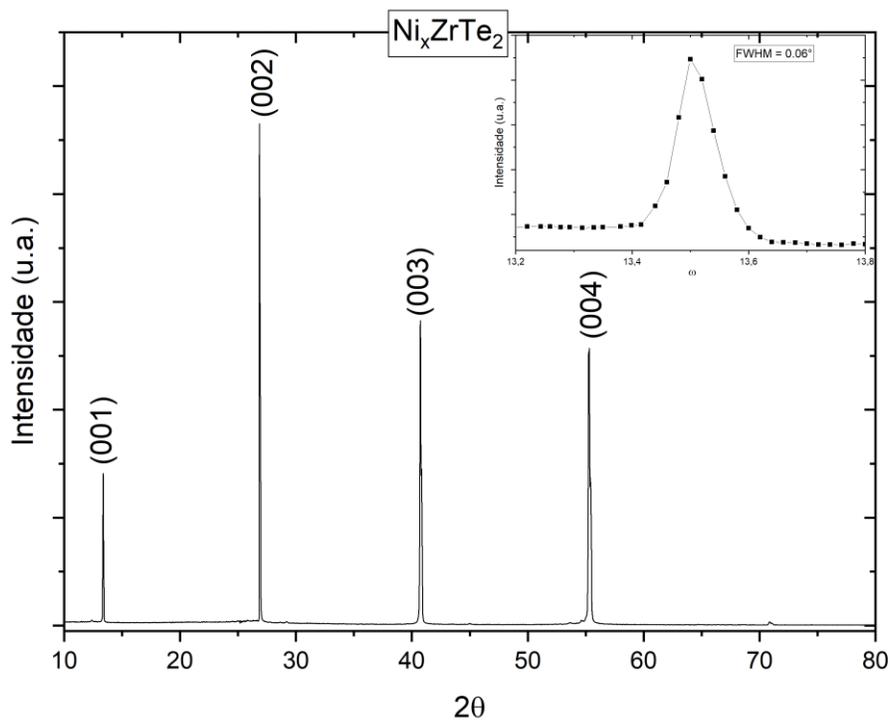

*Figure 4 – θ-2θ X-ray diffraction of a $Ni_{0.05}ZrTe_2$ crystal. The X-ray beam is directed to the flat surface of the crystal (ab-plane). The (00l) are the only observed reflections, indicating that the flat facets correspond to the ab-plane of the CdI₂-type structure. The inset displays a rocking curve centered on the (001) reflection. The narrow-spread is indicative of excellent crystallinity.*

Fig. 4 shows clearly that only the (*00l*) planes can be observed when the X-ray beam is incident on the flat surfaces. Given that the $T_xR$Te₂ and $L_xR$Te₂ crystals form with the hexagonal CdI₂ structure, the larger flat facets must correspond to the ab-plane. A rocking curve centered on the (001) reflection shows a narrow FWHM of 0.06°, consistent with a high level of crystallinity.

The SEM with EDS analysis show that the stoichiometric ratio Zr:Te is preserved at 1:2 in the growth process, similarly to the growths of TiTe₂ and HfTe₂. The concentration of intercalating species varies, though typically it stays within the 0.02-0.05 range. These values were determined by EDS and ICP, which typically yielded consistent values. An analysis of the pellet composition after the growth indicated a great variation from the initial stoichiometry, suggesting that the growth was halted when the composition of the pellets deviated from the initial conditions past a certain limit. The crystal growth is interrupted when a thermodynamic equilibrium is reached between the grown crystals and the remaining pellet, therefore halting mass transport.

### Conclusion

In summary, this work describes the growth of RTe$_2$ as well as intercalated $T_xR$Te$_2$ and $L_xR$Te$_2$ single crystals by means of an isothermal CVT method, using iodine as transport agent. The optimized concentration of iodine for the larger crystal size yields was $\approx$ 1.3 g/cm$^3$, resulting in crystals as large as 15 x 8 x 0.2 mm$^3$. The thermodynamic driving force for growth is the small difference in chemical potential created by iodine vaporization of the surface of the grains of the polycrystalline precursor. In contrast with the typical CVT growth, which is driven by temperature gradients, compounds having a large homogeneity range can be transported effectively from polycrystalline to monocrystalline state under isothermal conditions, through the chemical potential gradient created between the sample and the gaseous phase. Finally, the single crystals of this study reached their larger sizes after about 7 days of isothermal growth, and showed excellent crystallographic quality.

### Acknowledgments

The authors would like to thank the funding agencies: FAPESP (2018/08819-2, 2019/14359-7), CNPq (311394/2021-3), FAPEMIG (APQ-02861-21) and CAPES (88887.595960/2020-00).